\documentclass[aps,pra,a4paper,superscriptaddress,twocolumn,showpacs,showkeys,longbibliography,floatfix]{revtex4-1}

\usepackage{amsmath}
\usepackage{natbib}
\usepackage[utf8]{inputenc}
\usepackage[sans]{dsfont}
\usepackage{marvosym}
\usepackage{graphicx}

\DeclareMathOperator{\Tr}{Tr}
\DeclareMathOperator{\diff}{d}
\newcommand{\EXP}[1]{\ensuremath{\mathrm{e}^{#1}}}
\newcommand{\eg}{{\it{e.g.}}}
\newcommand{\ie}{{\it{i.e.}}}
\newcommand{\etal}{{\it{et al.~}}}
\newcommand{\refsec}[1]{Sec.~\ref{#1}}
\newcommand{\refeq}[1]{Eq.~(\ref{#1})}
\newcommand{\reffig}[1]{Fig.~\ref{#1}}

\begin{document}


    \title{Generation of cluster states in optomechanical quantum systems}
    \date{August 05, 2015}

    \author{Oussama Houhou}
    \email[E-mail: ]{o.houhou@qub.ac.uk}
    \affiliation{Laboratoire de Physique Math\'ematique et Subatomique (LPMS), Universit\'e de Constantine 1, Constantine, Algeria}
    \affiliation{Centre for Theoretical Atomic, Molecular and Optical Physics, School of Mathematics and Physics, Queen’s University, Belfast BT7 1NN, United Kingdom}

    \author{Habib Aissaoui}
    \email[E-mail: ]{aissaoui\_h@yahoo.com}
    \affiliation{Laboratoire de Physique Math\'ematique et Subatomique (LPMS), Universit\'e de Constantine 1, Constantine, Algeria}
    
        \author{Alessandro Ferraro}
    \email[E-mail: ]{a.ferraro@qub.ac.uk}
    \affiliation{Centre for Theoretical Atomic, Molecular and Optical Physics, School of Mathematics and Physics, Queen’s University, Belfast BT7 1NN, United Kingdom}
    
   \begin{abstract}
We consider an optomechanical quantum system composed of a single cavity mode interacting with $N$ mechanical resonators. We propose a scheme for generating continuous-variable graph states of arbitrary size and shape, including the so-called cluster states for universal quantum computation. The main feature of this scheme is that, differently from previous approaches, the graph states are hosted in the mechanical degrees of freedom rather than in the radiative ones. Specifically, via a $2N$-tone drive, we engineer a linear Hamiltonian which is instrumental to dissipatively drive the system to the desired target state. The robustness of this scheme is assessed against finite interaction times and mechanical noise, confirming it as a valuable approach towards quantum state engineering for continuous-variable computation in a solid-state platform.
 \end{abstract}


    \maketitle



\section{Introduction}

Recent experimental advances have shown that various types of mechanical oscillators can operate deeply in the quantum regime \cite{meystre2013short,aspelmeyer2014cavity,rogers2014hybrid}, promoting these systems to interesting candidates for quantum technologies. In particular, different cooling techniques have succeeded in bringing these oscillators close to their ground state \cite{Connell:2010,teufel:2011,chan:2011,safavi:2012,brahms2012optical,verhagen2012quantum}, whereas the ability to realize a coherent radiation-pressure interaction between electromagnetic and mechanical degrees of freedom has allowed for the realization of genuine quantum processes, such as quantum state transfer  \cite{palomaki2013coherent} and the generation of squeezing \cite{brooks2012non,purdy2013strong,safavi2013squeezed} and entanglement \cite{palomaki2013entangling}. These achievements, together with the possibility to scale up the number of involved oscillators \cite{Bhattacharya2008,hartmann2008steady,xuereb2012strong,chang2011slowing}, pave the way for more advanced quantum information applications such as the engineering of quantum dissipation \cite{tomadin2012reservoir}, quantum many-body simulators \cite{ludwig2013quantum,seok2013multimode,xuereb2014reconfigurable}, and quantum information processing in general \cite{stannigel2012optomechanical,schmidt2012optomechanical}. 

In this context, Schimidt \etal \cite{schmidt2012optomechanical} have proposed a platform, based on the linearized radiation-pressure interaction, to implement general Gaussian operations \cite{Ferraro05,Weedbrook:12,adesso2014continuous} between multiple mechanical oscillators. The implementation of such a platform would represent a first step towards the realization of the circuit model of universal quantum computation over continuous variables \cite{lloyd98computation,Braunstein:05}. However, a valid alternative approach to quantum computation is constituted by the so called measurement-based model \cite{Raussendorf:01}. The latter allows to perform general processing of quantum information over continuous variables \cite{menicucci2006universal} provided a massively entangled state --- dubbed \textit{cluster state} --- is used as a resource and additional measurements are locally performed over its constituents. Despite the limitations of finite squeezing \cite{ohliger2010limitations,ohliger2012efficient}, this approach has been proven to be fault tolerant \cite{menicucci2014fault} and, as a matter of fact, much effort has been devoted towards the generation of cluster states of light, culminating with the experimental realization of states composed of a high number of modes (up to $10,000$) both in the time \cite{yokoyama2013ultra} and frequency \cite{chen2014experimental,roslund2014wavelength} domain.
However, specific schemes for the generation of cluster states involving massive degrees of freedoms, rather than radiative ones, are still lacking --- despite some theoretical framework generically suitable for their implementation has been proposed \cite{aolita2011gapped,Menicucci:11}. The main advantage of this type of cluster states is that, being hosted in stationary or solid-state based architectures, they offer a promising path towards integrated and scalable quantum technologies.   

In order to bridge this gap, the aim of the present work is to introduce a scheme to generate continuous-variable cluster states of mechanical oscillators. We propose a scheme for generating graph states \footnote{The reader should be aware of various uses of the terms cluster and graph state in the context of continuous-variable quantum computation \cite{Weedbrook:12}. Here, we use the convention of referring to a cluster state as one which is universal for measurement-based computation (e.g., a square lattice); while a graph state could be a state represented by an arbitrary graph. Also, notice that we always refer to Gaussian pure states.} of arbitrary size and shape (including cluster states) whose nodes are embodied by the mechanical modes of an optomechanical system. The graph states are obtained by properly engineering both the Hamiltonian and the dissipative dynamics of the radiation degrees of freedom. Specifically, the method we use to engineer the desired Hamiltonian is based on multi-tone external driving, adapting and generalizing previous approaches \cite{clerk2013squeezing,tan2013achieving,tan2013dissipation,woolley2014two,abdi2015entangling,li2015generation} so that the required sidebands could be independently excited. In order to drive dissipatively the system to the graph states, we use a theoretical framework --- introduced in Ref.~\cite{Yamamoto:13} --- that adapts quantum dissipation engineering to Gaussian continuous-variable systems. The merit of our scheme is that one can generate arbitrary graph states only by driving the optomechanical system with a sequence of tunable pulses. 

The paper is organized as follows. In \refsec{sec:system} we introduce the system under consideration and derive the tunable linearized Hamiltonian that will be instrumental for the graph-state generation. In \refsec{sec:cluster-state} we will introduce the state generation protocol for the case of a generic graph state and in the absence of mechanical noise. We illustrate the action of the protocol via considering two specific examples. The detrimental effect of mechanical noise will be considered in detail in \refsec{sec:mech-noise}, confirming the robustness of the present protocol for low noise. A brief discussion about the experimental feasibility of our scheme is given in \refsec{sec:feasibility}. Finally, \refsec{sec:conclusions} will close the paper with some concluding remarks.


\section{Tunable linearized Hamiltonian of the system}\label{sec:system}
\begin{figure}[h]
\centering
\includegraphics[width=\columnwidth]{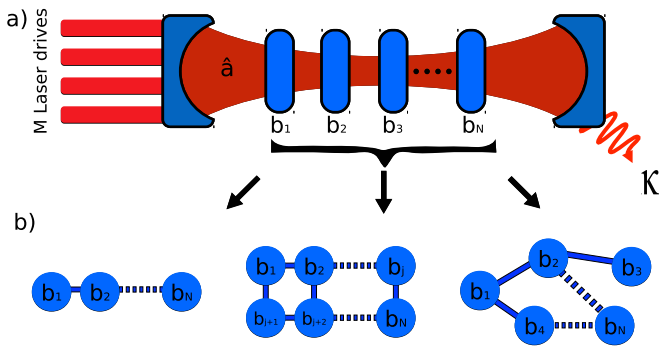}
\caption{(Color online) (a) : An optomechanical system consisting of one optical cavity mode $a$ coupled to $N$ non-interacting mechanical resonators $b_1,\dots,b_N$. The cavity dissipates with a damping rate $\kappa$, and it is driven by $M$ classical laser fields. (b) : The state of the mechanical resonators can be prepared in different graph state geometries, \textit{e.g.}, from left to right, a linear, a dual-rail, and a generic graph state. 
\label{fig:optomech-system}}
\end{figure}

	Consider an optomechanical system consisting of $N$ non-interacting mechanical resonators and a single-mode optical cavity driven by $M$ classical laser fields [$M$-tone drive, see \reffig{fig:optomech-system}-(a)].  This configuration extends and adapts to our purposes some approaches already considered in the literature in order to generate single-mode squeezing \cite{clerk2013squeezing} and entanglement among two mechanical oscillators \cite{tan2013achieving, tan2013dissipation, woolley2014two, abdi2015entangling, li2015generation}. In this section, we derive the family of tunable Hamiltonians that we will use later on to generate our target states. 

The Hamiltonian of the system described in \reffig{fig:optomech-system} is given by \cite{meystre2013short,aspelmeyer2014cavity}
	\begin{equation}
		\mathcal{H}=\omega_c a^\dagger a+\sum_{j=1}^N\left[\Omega_j b_j^\dagger b_j+g_j a^\dagger a(b_j^\dagger+b_j)\right]+\epsilon(t) a^\dagger+\epsilon^*(t)a
	\end{equation}
	where $a$ and $\omega_c$ ($b_j$ and $\Omega_j$) are the annihilation operator and frequency of the cavity ($j^{\mathrm{th}}$ mechanical-resonator) mode respectively, whereas $g_j$ is the one photon coupling to the $j^\mathrm{th}$~mechanical resonator mode. The driving laser fields $\epsilon(t)$ are given by
	\begin{equation}
		\epsilon(t)=\sum_{k=1}^M\epsilon_k\EXP{-i\omega_k t}\EXP{i\Phi_k}
	\end{equation}
	where $\omega_k$, $\epsilon_k$ and $\Phi_k$ are respectively the frequency, intensity and phase of the $k^\mathrm{th}$~laser field.

	The equation of motion for the optical and mechanical modes are written as:
	\begin{eqnarray}
		\dot{a}	&=&	-i\omega_c a-ia\sum_{j=1}^Ng_j(b_j^\dagger+b_j)-i\epsilon(t) \nonumber \\ & & -\frac{\kappa}{2}a-\sqrt{\kappa}a_{in}\\
		\dot{b_j}	&=&		-i\Omega_j b_j-ig_j a^\dagger a\label{eqn:equ-motion-mech}
	\end{eqnarray}
	where $a_{in}$ is the input noise operator for the cavity mode with decay rate $\kappa$.  In Eq.~(\ref{eqn:equ-motion-mech}) we have omitted the terms describing the mechanical noise, assuming that the cavity damping process dominates all other noisy processes. This requirement will be relaxed in \refsec{sec:mech-noise}, where we will consider the robustness of our scheme against mechanical noise.

	Now, we linearise the equations of motion by considering the following:
	\begin{eqnarray}
		g_j &&\ll \omega_c,\Omega_j\\
		a &&\rightarrow a+\alpha\label{eqn:split-a}\\
		b_j &&\rightarrow b_j+\beta_j\label{eqn:split-b}
	\end{eqnarray}
	where, in Eqs.~(\ref{eqn:split-a}) and~(\ref{eqn:split-b}), we split the optical (mechanical) field into classical part $\alpha$ ($\beta_j$) and a quantum noise term $a$ ($b_j$). Neglecting the second-order terms in  $a$ and $b_j$, the equations of motion become:
	\begin{eqnarray}
		\dot{a}		&\approx&		-i\omega_c a-ia\sum_{j=1}^N g_j(\beta_j^*+\beta_j) \nonumber \\		
	& &-i\alpha\sum_{j=1}^Ng_j(b_j^\dagger+b_j)-\frac{\kappa}{2}a-\sqrt{\kappa}a_{in}\\
		\dot{\alpha}	&\approx&		-i\omega_c \alpha-i\alpha\sum_{j=1}^Ng_j(\beta_j^*+\beta_j)-i\epsilon(t)-\frac{\kappa}{2}\alpha\label{eqn:alpha-diff}\\
		\dot{b_j}		&\approx&		-i\Omega_j b_j-ig_j(\alpha a^\dagger+\alpha^*a)\\
		\dot{\beta_j}	&\approx&		-i\Omega_j \beta_j-ig_j|\alpha|^2
	\end{eqnarray}
	and the linearised Hamiltonian reads:
	\begin{equation}
		\mathcal{H}=\omega'_c a^\dagger a+\sum_{j=1}^N\Omega_j b_j^\dagger b_j+\sum_{j=1}^N g_j(\alpha a^\dagger+\alpha^* a)(b_j^\dagger+b_j)
	\end{equation}
	with $\omega'_c=\omega_c+\sum_{j=1}^N g_j(\beta_j^*+\beta_j)$. For $|\beta_j|g_j\ll\omega_c$ we have $\omega'_c\approx\omega_c$.

	From Eq.~(\ref{eqn:alpha-diff}), the parameter $\alpha$ is approximately given by
	\begin{equation}
		\alpha=\sum_{k=1}^M\frac{-i\epsilon_k}{\kappa/2+i(\omega_c-\omega_k)}\EXP{-i\omega_k t}\EXP{i\Phi_k}\;.
	\end{equation}
By defining $\alpha_k\EXP{i\phi_k}:=\frac{-i\epsilon_k}{\kappa/2+i(\omega_c-\omega_k)}\EXP{i\Phi_k}$ with $\alpha_k\ge 0$ and $\phi_k\in\mathds{R}$,  the linearised Hamiltonian becomes:
	\begin{eqnarray}
		\mathcal{H} &=& \omega_c a^\dagger a+\sum_{j=1}^N\Omega_j b_j^\dagger b_j \nonumber \\
		&+& \left[\sum_{j=1}^N\sum_{k=1}^M\alpha_k g_j\EXP{i\phi_k}\EXP{-i\omega_k t}a^\dagger(b_j^\dagger+b_j)+\mathrm{H.c.}\right].\label{eqn:optomech-hamilt}
	\end{eqnarray}

In order to eliminate the explicit time dependence of the Hamiltonian we first go to the interaction picture
	\begin{eqnarray}
	H &=& a^\dagger\sum_{j=1}^N\sum_{k=1}^M\alpha_k g_j\EXP{i\phi_k}\EXP{i(\omega_c-\omega_k)t}\left(\EXP{i\Omega_j t}b_j^\dagger + \EXP{-i\Omega_j t}b_j\right) \nonumber \\
& & +\mathrm{H.c.}\label{eqn:hamiltonian-before-rwa}
	\end{eqnarray}	
We then consider for each mechanical resonator~$j$, two laser drives with frequencies $\omega_j^\pm=\omega_c\pm\Omega_j$, amplitudes $\alpha_j^\pm$, and phases $\phi_j^\pm$. Assuming that the mechanical frequencies do not overlap and that we are in the weak coupling regime
	\begin{equation}\label{eqn:RWA-condition}
		\alpha_j^\pm g_j\ll\Omega_j\;,
	\end{equation}
we can invoke the rotating wave approximation, finally obtaining the following Hamiltonian:
	\begin{equation}
		H=a^\dagger\sum_{j=1}^N g_j\left(\alpha_j^+\EXP{i\phi_j^+}b_j^\dagger
		+\alpha_j^-\EXP{i\phi_j^-}b_j\right)
		+\mathrm{H.c.}\label{eqn:hamiltonian}
	\end{equation}
	
The main feature of the Hamiltonian above is that it includes both ``beam-splitter'' and ``two-mode squeezing'' interactions that can be independently tuned for each mechanical oscillator. In particular, we will see in the next Section that, by controlling the amplitude and relative phase of each driving laser, one can conveniently tune the parameters $\alpha_j^\pm$ and $\phi_j^\pm$ to generate the desired graph state.


\section{Graph-state generation protocol}\label{sec:cluster-state}
As said, our aim is to devise a scheme to generate arbitrary graph states of the mechanical degrees of freedom for the system described in \refsec{sec:system} [see \reffig{fig:optomech-system}-(b)]. We will show now how this goal can be achieved by exploiting as a resource the dissipation of the radiation mode, together with a suitably tuned sequence of laser pulses. For the sake of clarity, in this Section we will neglect the dissipation of the mechanical degrees of freedom, which will be taken into account in  \refsec{sec:mech-noise}.
	
In general, the dissipation into an inaccessible environment causes the loss of quantum coherence, driving the system of interest into a state void of genuine quantum features. However, dissipation can indeed turn into a resource that does stabilise quantum coherence, provided the dissipative dynamics is properly engineered \cite{poyatos1996quantum,carvalho2001decoherence,plenio2002entangled,diehl2008quantum,verstraete2009quantum}. In the context of continuous-variable systems,  dissipation engineering has been considered for the purpose of entanglement generation, both at a general theoretical level \cite{nurdin2009network,Yamamoto:12,Yamamoto:13,ma2014preparation} and in experiments involving atomic ensembles  \cite{krauter2011entanglement,muschik2011dissipatively,muschik2012robust}. Concerning the specific setting of optomechanics, recent theoretical proposals have shown how dissipation engineering can allow to achieve strong steady-state squeezing \cite{clerk2013squeezing} and two-mode entanglement in three- and four-mode systems \cite{tan2013achieving,wang2013reservoir,tan2013dissipation,woolley2014two,chen2015dissipation,abdi2015entangling,li2015generation}. Here, we generalize these approaches and propose a scheme to achieve an arbitrary graph state at the steady state. 

The dynamics of the system is described by the master equation
	\begin{equation}
		\frac{\diff\rho}{\diff t}=-i[H,\rho]+\mathcal{L}\label{eqn:master-equation}
	\end{equation}
	with $\rho$ is the system's density matrix, $H$ is given by Eq.~(\ref{eqn:hamiltonian}), and $\mathcal{L}$ accounts for the dissipation processes. Assuming that the cavity damping process dominates all other dissipative processes, $\mathcal{L}$ reads as follows:
	\begin{equation}\label{eqn:cavity-dissipation}
		\mathcal{L}=\kappa(a\rho a^\dagger-\frac{1}{2} a^\dagger a\rho-\frac{1}{2}\rho a^\dagger a).
	\end{equation}
One can show that the system above admits a unique pure Gaussian steady-state only if just a single mechanical mode is involved (see Appendix~\ref{sec:appendix-uniqueness}). This obstacle can be circumvented, in the case of many mechanical resonators, using the mechanism of \textit{Hamiltonian switching} proposed by Li {\em et al.} \cite{Li:09} and generalised by Ikeda and Yamamoto \cite{Yamamoto:13}. The key point of this approach is that only one suitable collective mode is in fact coupled at each switching step. This is achieved by dividing the dynamics in as many steps as the number $N$ of (mechanical) modes involved. At each step, the interaction Hamiltonian has to be properly engineered in order for the optical dissipation to cool a specific collective mechanical mode. Once the steady  state is achieved, another collective mode is cooled by switching to a new Hamiltonian (with the remaining collective modes uncoupled). The main difficulty of this scheme is to design a system that sustains the set of Hamiltonians required to generate a certain desired state, in our case an arbitrary Gaussian graph state.

\begin{table*}[t]
\begin{ruledtabular}
\begin{tabular}{c|cccccccc}
			Step &	$\alpha_1^- $ & $\alpha_2^-$ & $\alpha_3^-$ & $\alpha_4^-$ & 
					$\phi_1^- $ & $\phi_2^-$ & $\phi_3^-$ & $\phi_4^-$\\
\hline
1 & $\frac{\sqrt{2(5+\sqrt{5})}}{5}$ & $\frac{\sqrt{5+2\sqrt{5}}}{5}$ & $\frac{\sqrt{5-2\sqrt{5}}}{5}$ & $\frac{\sqrt{5+2\sqrt{5}}}{5}$ & $3\pi/2$ & $\pi$ & $\pi/2$ & 0\\
2 & $\frac{\sqrt{5+2\sqrt{5}}}{5}$ & $\frac{\sqrt{5+2\sqrt{5}}}{5}$ & $\frac{\sqrt{2(5-\sqrt{5})}}{5}$ & $\frac{\sqrt{5-2\sqrt{5}}}{5}$ & $\pi$ & $3\pi/2$ & $\pi$ & $\pi/2$\\
3 & $\frac{\sqrt{5-2\sqrt{5}}}{5}$ & $\frac{\sqrt{2(5-\sqrt{5})}}{5}$ & $\frac{\sqrt{5+2\sqrt{5}}}{5}$ & $\frac{\sqrt{5+2\sqrt{5}}}{5}$ & $\pi/2$ & $\pi$ & $3\pi/2$ & $\pi$\\
4 & $\frac{\sqrt{5+2\sqrt{5}}}{5}$ & $\frac{\sqrt{5+2\sqrt{5}}}{5}$ & $\frac{\sqrt{2(5-\sqrt{5})}}{5}$ & $\frac{\sqrt{5-2\sqrt{5}}}{5}$ & 0 & $\pi/2$ & $\pi$ & $3\pi/2$\\
\end{tabular}
\end{ruledtabular}
\caption{Assignement of the parameters in \refeq{eqn:hamiltonian} for each of the four steps of the Hamiltonian switching procedure that allows to generate a four-mode linear graph state.}\label{tab:lin_cluster}
\end{table*}

\begin{table*}[t]
\begin{ruledtabular}
\begin{tabular}{c|cccccccc}
Step & $\alpha_1^- $ & $\alpha_2^-$ & $\alpha_3^-$ & $\alpha_4^-$ & 
 $\phi_1^- $ & $\phi_2^-$ & $\phi_3^-$ & $\phi_4^-$\\
\hline
1 & $\frac{5+\sqrt{5}}{10}$ & $\frac{1}{\sqrt{5}}$ & $\frac{-5+\sqrt{5}}{10}$ & $\frac{1}{\sqrt{5}}$ & $3\pi/2$ & $\pi$ & $\pi/2$ & $\pi$\\
2 & $\frac{1}{\sqrt{5}}$ & $\frac{5+\sqrt{5}}{10}$ & $\frac{1}{\sqrt{5}}$ & $\frac{-5+\sqrt{5}}{10}$ & $\pi$ & $3\pi/2$ & $\pi$ & $\pi/2$\\
3 & $\frac{-5+\sqrt{5}}{10}$ & $\frac{1}{\sqrt{5}}$ & $\frac{5+\sqrt{5}}{10}$ & $\frac{1}{\sqrt{5}}$ & $\pi/2$ & $\pi$ & $3\pi/2$ & $\pi$\\
4 & $\frac{1}{\sqrt{5}}$ & $\frac{-5+\sqrt{5}}{10}$ & $\frac{1}{\sqrt{5}}$ & $\frac{5+\sqrt{5}}{10}$ & $\pi$ & $\pi/2$ & $\pi$ & $3\pi/2$\\
\end{tabular}
\end{ruledtabular}
\caption{Assignement of the parameters in \refeq{eqn:hamiltonian} for each of the four steps of the Hamiltonian switching procedure that allows to generate a four-mode square graph.}\label{tab:square_cluster}
\end{table*}

More specifically, consider a generic quadratic unitary transformation $U$ \citep{Ferraro05,Weedbrook:12,adesso2014continuous} acting as $c=Ub$ on the mechanical modes $b=(b_1,\ldots,b_N)^T$, where $c~=~(c_1,\ldots,c_N)^T$ defines a set of collective modes. As said, we perform an $N$-step transformation, where in the $k^{\rm{th}}$ step we set the driving laser's amplitudes and phases in \refeq{eqn:hamiltonian} as follows:
	\begin{eqnarray}
		\alpha_j^-	&=&		\frac{\beta}{g_j}|U_{kj}|\label{eqn:Ham_parameters}\\
		\alpha_j^+	&=&		r\alpha_j^-\label{eqn:alpha-plus}\\
		\phi_j^-		&=&		-\phi_j^+=\mathrm{arg}(U_{kj})\label{eqn:phi_plus}
	\end{eqnarray}
where $\beta>0$ and $0\le r<1$. With this settings \refeq{eqn:hamiltonian} becomes:
	\begin{equation}\label{eqn:hamiltonian-step-k}
		H\equiv H^{(k)}=\beta a^\dagger(c_k+r c_k^\dagger)+\mathrm{H.c.}
	\end{equation}
One can show (see Appendix~\ref{sec:appendix-uniqueness}) that the above Hamiltonian, with the help of the cavity dissipation in Eq.~(\ref{eqn:cavity-dissipation}), generates at the steady state a single-mode squeezed state for the collective mode $c_k$. After applying all the $N$ steps of the transformation, we obtain $N$~single-mode squeezed state relative to the modes $c_1,\ldots,c_N$. In terms of the mechanical modes $b_1,\ldots,b_N$, the latter corresponds to a Gaussian state with zero mean. Let us introduce the canonical position and momentum operators for each mechanical mode $q_j=(b_j^\dagger+b_j)/\sqrt2$ and $p_j=i(b_j^\dagger-b_j)/\sqrt2$, respectively, the vector $R=(q_1,\dots,q_N,p_1,\dots,p_N)$, and the covariance matrix $V$ whose elements are $[V]_{kl}=\langle R_k R_l+R_l R_k\rangle /2$. Then the covariance matrix of the mechanical Gaussian state reads:
	\begin{equation}
		V=\frac{1}{2}S^T
		\left(
		\begin{array}{l@{\quad}l}
			\EXP{-2\xi}\mathds{1}^{N\times N}	&	0^{N\times N}\\
			0^{N\times N}						&	\EXP{2\xi}\mathds{1}^{N\times N}
		\end{array}
		\right)
		S
		\label{eq:covariance_mat}
	\end{equation}
where $\xi=\tanh^{-1}r$ and $S$ is the sympletic transformation corresponding to the quadratic unitary transformation $U$ \cite{Ferraro05,Weedbrook:12,adesso2014continuous}. The quantity $\EXP{-2\xi}$ is the squeezing parameter and the state given by the covariance matrix (\ref{eq:covariance_mat}) has a level of squeezing of $10\log_{10}[\EXP{2\xi}]$dB. Ikeda and Yamamoto \cite{Yamamoto:13} showed that the unitary transformation $U$ that generates an arbitrary Gaussian graph state is obtained from the polar decomposition of the matrix $-(i\mathds{1}_N+A)=RU$ where $A$ is the adjacency matrix describing the graph state that one wants to generate \cite{zhang2006continuous,gu2009quantum}, and $R$ is $N\times N$ real matrix. 

\subsection{Examples}

In the remaining part of this section, we give explicit examples demonstrating how to generate a linear and a square four-mode graph states using the foregoing scheme. We have chosen four modes only for the sake of clarity but, in principle, the size and shape of graph state that can be engineered is arbitrary. For simplicity we set $g_j=g$ for all $j=1,\ldots,4$, and all the amplitudes are given in  units of $\beta/g$.
	
Let us first consider a four-mode equally weighted linear graph state. This state is a basic building block in measurement-based quantum computation since it allows to implement an arbitrary single-mode Bogoliubov transformation via local measurements \cite{ukai2010universal,ukai2011demonstration}. The adjacency matrix is given by:
		$$A=
			\left(
			\begin{array}{c@{\quad}c@{\quad}c@{\quad}c@{\quad}}
				0	&	1	&	0	&	0\\
				1	&	0	&	1	&	0\\
				0	&	1	&	0	&	1\\
				0	&	0	&	1	&	0
			\end{array}
			\right)
		$$
from which, as explained above, we obtain the corresponding unitary matrix $U$ via the polar decomposition of the matrix $-(i\mathds{1}_4+A)$. The graph state is then generated in four steps. In each step the parameters of the driving laser fields are tuned to the values obtained from Eqs.~(\ref{eqn:Ham_parameters}) and (\ref{eqn:phi_plus}) and indicated in Tab.~\ref{tab:lin_cluster} [$\alpha_j^+$ is always given by \refeq{eqn:alpha-plus} and $\phi_j^+=-\phi_j^-$].

As a second example, we consider a four-mode equally weighted square graph state. The relevance of the latter stems from the fact that it allows to perform two-mode quantum operations and, in addition, it can be used in a redundant encoding scheme for error filtration \cite{van2007building}. The corresponding adjacency matrix is :
		$$A=
			\left(
			\begin{array}{c@{\quad}c@{\quad}c@{\quad}c@{\quad}}
				0	&	1	&	0	&	1\\
				1	&	0	&	1	&	0\\
				0	&	1	&	0	&	1\\
				1	&	0	&	1	&	0
			\end{array}
			\right).
		$$
As before, the graph state is generated in four steps, where in each step, the parameters of the laser fields are set according to the values indicated in Tab.~\ref{tab:square_cluster}.

\begin{figure}[h]
\centering
\includegraphics[width=\columnwidth]{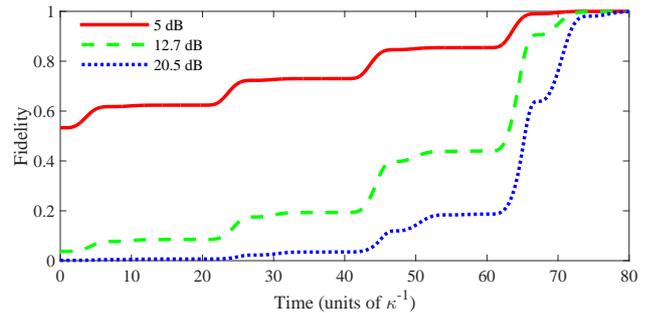}
\caption{\label{fig:fidelity-evolution} (Color online) Time evolution of the fidelity for a four-modes linear graph state with fixed switching time $t_{\rm s}=20\kappa^{-1}$. We have set for the $j^{th}$ oscillator the frequency $\Omega_j/ 2\pi= j~{\rm MHz}$.}
\end{figure}

\begin{figure}[h]
\centering
\includegraphics[width=\columnwidth]{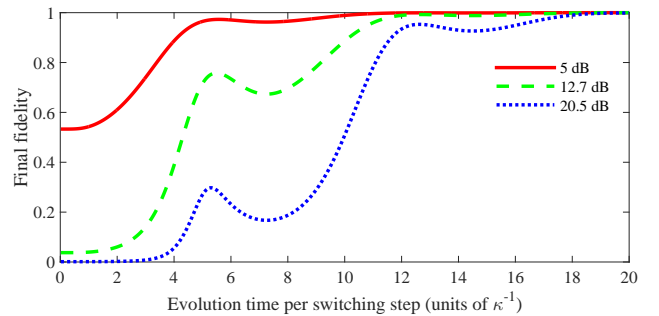}
\caption{\label{fig:fidelity-switch-time}(Color online) Fidelity of the steady state of the four-mode linear graph as a function of evolution time per switching step. The fidelity shown in this plot is that of the steady state of the mechanical oscillators after applying all the switching steps. We have set for the $j^{th}$ oscillator the frequency $\Omega_j/ 2\pi= j~{\rm MHz}$.}
\end{figure}

In principle, the results derived here are strictly valid only in the ideal case in which the steady state is reached at each step of the Hamiltonian switching scheme. In practice, the amount of time that can be devoted to each step is finite and one should assess the errors determined by the finite-time dynamics. The latter can be estimated using the Uhlmann fidelity between the ideal target state and the actual one, as a function of the switching time $t_{\rm s}$ (\ie, the time elapsing between one step and the other). Given two generic states $\rho_1$ and $\rho_2$, the  Uhlmann fidelity is defined as $F(\rho_1,\rho_2)=\left(\Tr\sqrt{\sqrt{\rho_1}\rho_2\sqrt{\rho_1}}\right)^2$ , which satisfies $0\le F(\rho_1,\rho_2)\le 1$ and $F(\rho_1,\rho_2)=1$ if and only if the states $\rho_1$ and $\rho_2$ are identical \cite{Uhlmann:76}. The latter can be conveniently expressed in terms of covariance matrices for Gaussian states, using the expressions provided in Refs.~\cite{Paraoanu:00,Spedalieri:13}. In \reffig{fig:fidelity-evolution} we plot the time evolution of the fidelity for a finite switching time $t_{\rm s}$ and for the case of a four-mode linear cluster state. We can see that the fidelity approaches its final value $F_{\rm gen}$ step after step. In \reffig{fig:fidelity-switch-time} we report instead  the final fidelity $F_{\rm gen}$ as a function of $t_{\rm s}$, for the same target state. We can see that, as expected, the fidelity of the state generated via Hamiltonian switching and dissipation engineering approaches the unit value for large enough switching times $t_{\rm s}$. In Appendix~\ref{sec:appendix-speed-steady-state}, we give the time scale to reach the steady state at each step of the switching scheme. In the numerical simulations shown in \reffig{fig:fidelity-evolution} and \reffig{fig:fidelity-switch-time}, we have chosen $\beta=\kappa/(4\sqrt{1-r^2})$, so that the system reaches the steady state at each step in minimal time (see Appendix~\ref{sec:appendix-speed-steady-state}) and with minimum driving power [the driving power determines the value of $\alpha_j^\pm$, which are related to $\beta$ via \refeq{eqn:Ham_parameters}]. With this choice of $\beta$, using \refeq{eqn:Ham_parameters} and considering the condition~(\ref{eqn:RWA-condition}), the cavity decay rate must satisfy the following condition:
	\begin{equation}\label{eqn:kappa-condition}
		\kappa\ll 4\sqrt{1-r^2}\min_{k,j=1,\ldots,N}\frac{\Omega_j}{|U_{kj}|}\ .
	\end{equation}
The latter implies that \textit{(i)} the scheme here introduced works in the resolved-sideband regime and \textit{(ii)} the higher the squeezing the deeper into the good cavity regime one has to operate. 

\section{Robustness of the graph-state generation against mechanical noise} \label{sec:mech-noise}

\begin{figure*}[t]
\centering
\includegraphics[width=\textwidth,height=9cm]{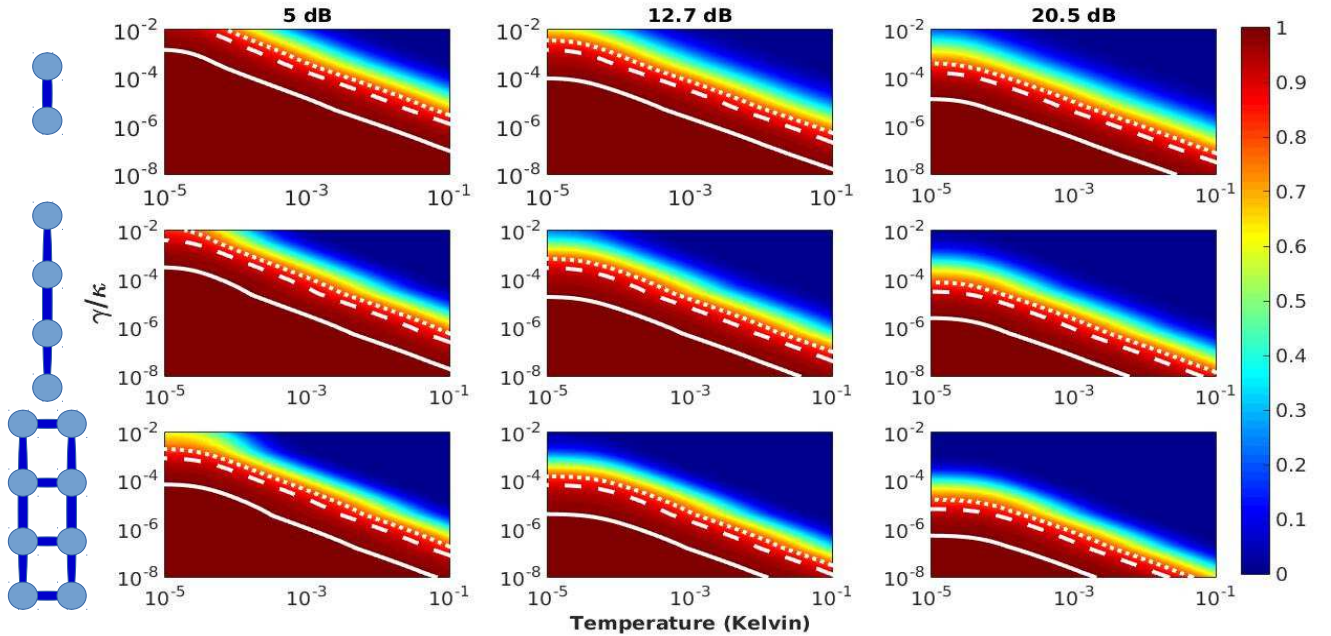}
\caption{\label{fig:fidelity-temp-damp}(Color online) Contour plots of the fidelity between the target state and the state generated using the protocol described in the text. The fidelity is shown as a function of the temperature of the mechanical bath (horizontal axis) and the mechanical damping rate (vertical axis) for different target graph states and levels of squeezing. Each mechanical oscillator has a frequency $\Omega_j/ 2\pi= j~\mathrm{MHz}$ ($j=1,\ldots,N$, with $N=2,4,8$). The solid white lines correspond to a fidelity of $0.99$, the dashed lines to $0.90$, and the dotted lines to $0.80$. Each data point is taken for an optimal evolution time (see text).}
\end{figure*}

We have hitherto showed how to generate the target state without considering any mechanical dissipation. We now relax this condition and consider the mechanical resonators in contact with a thermal bath, addressing the detrimental effect of the latter to the scheme introduced above.

	In the master equation~(\ref{eqn:master-equation}), we include the following additional decoherence channels:
	\begin{eqnarray}
		\mathcal{L}_1	&=&	\sum_{j=1}^N\gamma_j(n_j+1)\left(b_j\rho b_j^\dagger-\frac{1}{2}b_j^\dagger b_j\rho-\frac{1}{2}\rho b_j^\dagger b_j\right)\\
		\mathcal{L}_2	&=&	\sum_{j=1}^N\gamma_j n_j\left(b_j^\dagger\rho b_j-\frac{1}{2}b_j b_j^\dagger\rho-\frac{1}{2}\rho b_j b_j^\dagger\right)
	\end{eqnarray}
	with $\gamma_j\ll\Omega_j$, where $\gamma_j$ and $n_j=\left(\exp{\frac{\hbar\Omega_j}{K_B T_j}}-1\right)^{-1}$ are the damping rate and the mean phonon number of the $j^\mathrm{th}$ mechanical bath respectively, with $T_j$ being the bath temperatures. Notice that we are assuming high-Q mechanical oscillators, hence we have considered the Markovian master equation above \cite{clerk2013squeezing, tan2013achieving, tan2013dissipation, woolley2014two} rather then the full non-Markovian Brownian noise \cite{giovannetti2001phase}.

	The thermal baths interact with the system during the preparation of the graph states, which results in a deviation of the final state from the target one. To quantify this deviation, we use again the Uhlmann fidelity. To see the effect of the thermal noise on our results of \refsec{sec:cluster-state}, we calculate the fidelity as a function of the mechanical damping rates and the bath temperatures. In our numerical simulations, we took the latter equal for each mechanical oscillator ($\gamma=\gamma_1=\ldots=\gamma_N$, $T=T_1=\ldots=T_N$) and set mechanical modes frequencies as $\Omega_j/ 2\pi= j~\mathrm{MHz}$ ($j=1,\ldots,N$). Figure~\ref{fig:fidelity-temp-damp} shows the fidelity for 2, 4 and 8-modes graph states as a function of $\gamma/\kappa$ and $T$. Each point of our simulation was obtained by searching for the optimal evolution time of the system in order to have maximum fidelity: very short evolution times do not allow to reach the target graph state, while very long times lead to dominant decoherence effects caused by the thermal bath. In other words, in the presence of mechanical noise, the best fidelity is not achieved at the steady state but at an intermediate evolution time typically larger than $[\Re\left(\frac{\kappa}{4}-\sqrt{(\frac{\kappa}{4})^2-\beta^2(1-r^2)}\right)]^{-1}$ (see Appendix~\ref{sec:appendix-speed-steady-state}) --- so that the desired target state can be achieved --- and small compared to $N$ times the inverse of the thermal decoherence rate \ie, smaller then $\min_{1\le j\le N}\frac{N \hbar\Omega_j}{\gamma_j K_B T}$, so that the mechanical noise is not too detrimental during the $N$ switching steps. 

One might wonder whether the fact that we are no longer considering the steady state constitutes an issue. This is in fact not the case in the context of cluster state quantum computation, where it is not necessary to build the entire cluster ahead of the computational task to perform, but rather it is possible to build and consume the cluster during the computation itself (see, \eg, Refs.~\cite{benjamin2009prospects,menicucci2010singleQND,roncaglia2011sequential}).

In \reffig{fig:fidelity-temp-damp} we focused on three paradigmatic graph states and squeezing levels. The first row shows results for the simplest graph, consisting of two modes only. Up to local symplectic operations that do not change its entanglement, the latter is equivalent to the usual two-mode squeezed state whose generation in optomechanical systems have been already addressed in previous literature \cite{tan2013achieving, wang2013reservoir, tan2013dissipation, woolley2014two, chen2015dissipation, abdi2015entangling, li2015generation}. The second row refers to a four-mode linear graph, which as said allows to implement a general single-mode symplectic transformation. The third row concerns an eight-mode {\em dual rail} graph, which allows instead to implement a generic two-mode symplectic transformation. The last two graphs thus encompass the necessary building blocks for universal multi-mode Gaussian unitaries. Each column in \reffig{fig:fidelity-temp-damp} refers to a specific squeezing level for the target state: from left to right we set $5$dB, $12.7$dB, and $21$dB. The first level of squeezing coincides with the one of the largest optical cluster state reported to date \citep{yokoyama2013ultra}, whereas $12.7$dB is the highest optical squeezing experimentally achieved with one optical mode \cite{schnabel2010squeezing}. The third squeezing level is a theoretical upper bound to the squeezing required to perform universal fault-tolerant quantum computation over continuous variables (see Ref.~\cite{menicucci2014fault}). As before, the plots of \reffig{fig:fidelity-temp-damp} were obtained by setting $\beta=\kappa/(4\sqrt{1-r^2})$ in order to reach the maximum fidelity in minimum time (see Appendix~\ref{sec:appendix-speed-steady-state} for more details). As said, this implies that the cavity decay rate must satisfy condition~(\ref{eqn:kappa-condition}).

From \reffig{fig:fidelity-temp-damp} we can see that, for a given squeezing and number of modes of the target state, the fidelity between the latter and the state obtained with our protocol decreases as the temperature $T$ and the coupling with the bath $\gamma$ increase. This is to be expected since our protocol will end up in a steady state that is far from a pure state for large mechanical noise. In general, one can also see in \reffig{fig:fidelity-temp-damp} that the region of high fidelity shrinks from the left panels to the right ones. This means that, for a given graph, the higher the squeezing of the target the less the mechanical noise that can be tolerated. The same behaviour can be see from top to bottom, implying that for a given target squeezing the larger the graph state the less the mechanical noise tolerated.

Let us notice here that the choice of the frequencies used in Figs.~\ref{fig:fidelity-temp-damp} and \ref{fig:fidelity-exp} is not unique. Indeed, our protocol works as well for any other choices of the oscillators frequencies, as long as they do not overlap and the rotating wave approximation can therefore be applied to the Hamiltonian in \refeq{eqn:hamiltonian-before-rwa}.


\section{Experimental feasibility}\label{sec:feasibility}

In view of the results of the previous Sections, let us now discuss some aspects regarding the experimental feasibility of the state generation scheme.

First, let us consider the attainability of the system Hamiltonian given in \refeq{eqn:hamiltonian}. As mentioned, various experiments have recently succeeded in realizing the weak optomechanical interaction that we have considered here \cite{meystre2013short, aspelmeyer2014cavity, rogers2014hybrid}. The main requirement that differentiates our scheme from the latter is that we consider, rather than only one mechanical oscillator,   multiple oscillators with non-overlapping frequencies. The first implementations of such systems have been reported recently \cite{lin2010coherent, massel2012multimode, shkarin2014optically}, thus providing a promising route towards the realization of small optomechanical networks. In addition, in quantum electromechanical systems, mechanically complaint membranes of slightly different geometry and size allow for the realization of mechanical oscillators with non overlapping frequencies --- as required in order for the rotating wave approximation adopted in \refeq{eqn:hamiltonian} to be valid. For example, the experiments described in Refs. \cite{wollman2015quantum,weinstein2014observation,teufel:2011,pirkkalainen2015squeezing} report a mechanical frequency of $3.6~\mathrm{MHz}$, $4~\mathrm{MHz}$, $10.56~\mathrm{MHz}$, and $13.03~\mathrm{MHz}$ respectively. This wide range of frequencies, obtained in compatible experimental set-ups, conveniently suits our proposal. More in general, the variety of mechanical frequencies realized in optomechanical systems (even within similar settings) suggests that the realization of small optomechanical networks with non-overlapping frequencies should be within reach.

\begin{figure}[h]
		\centering
		\includegraphics[width=0.9\columnwidth]{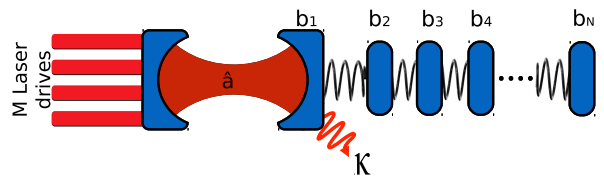}
		\caption{\label{fig:optomech-system2}(Color online) An optomechanical system consisting of one optical cavity mode and $N$ mechanical resonators modes. The cavity mode couples only to the first mechanical mode, and all the mechanical modes couple to each other via nearest-neighbour interaction.}
	\end{figure}

We should mention here that, despite the fact that our results are obtained for non-interacting non-degenerate mechanical resonators, the scheme here introduced can be easily extended to the case of interacting mechanical modes. In this case, the cavity mode could interact with one mechanical oscillator only (see \reffig{fig:optomech-system2}). After diagonalising the mechanical Hamiltonian, one obtains a set of non-interacting mechanical normal modes with non-overlapping frequencies, thus recovering the case considered here. The non-degeneracy of the normal modes could in principle be enforced by controlling the coupling between the mechanical modes.
	
\begin{figure}[t]
\centering
\includegraphics[width=\columnwidth]{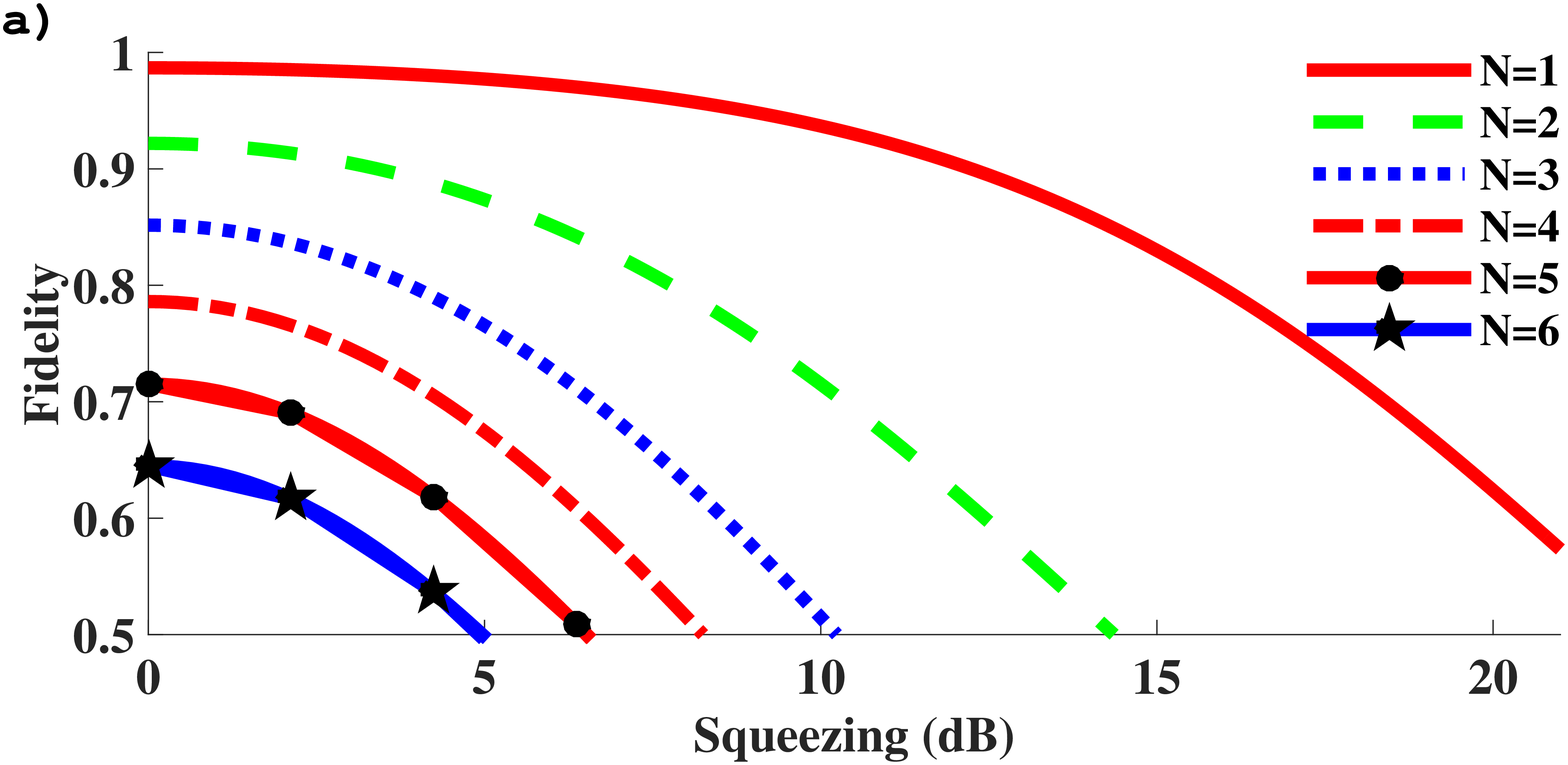}
\includegraphics[width=\columnwidth]{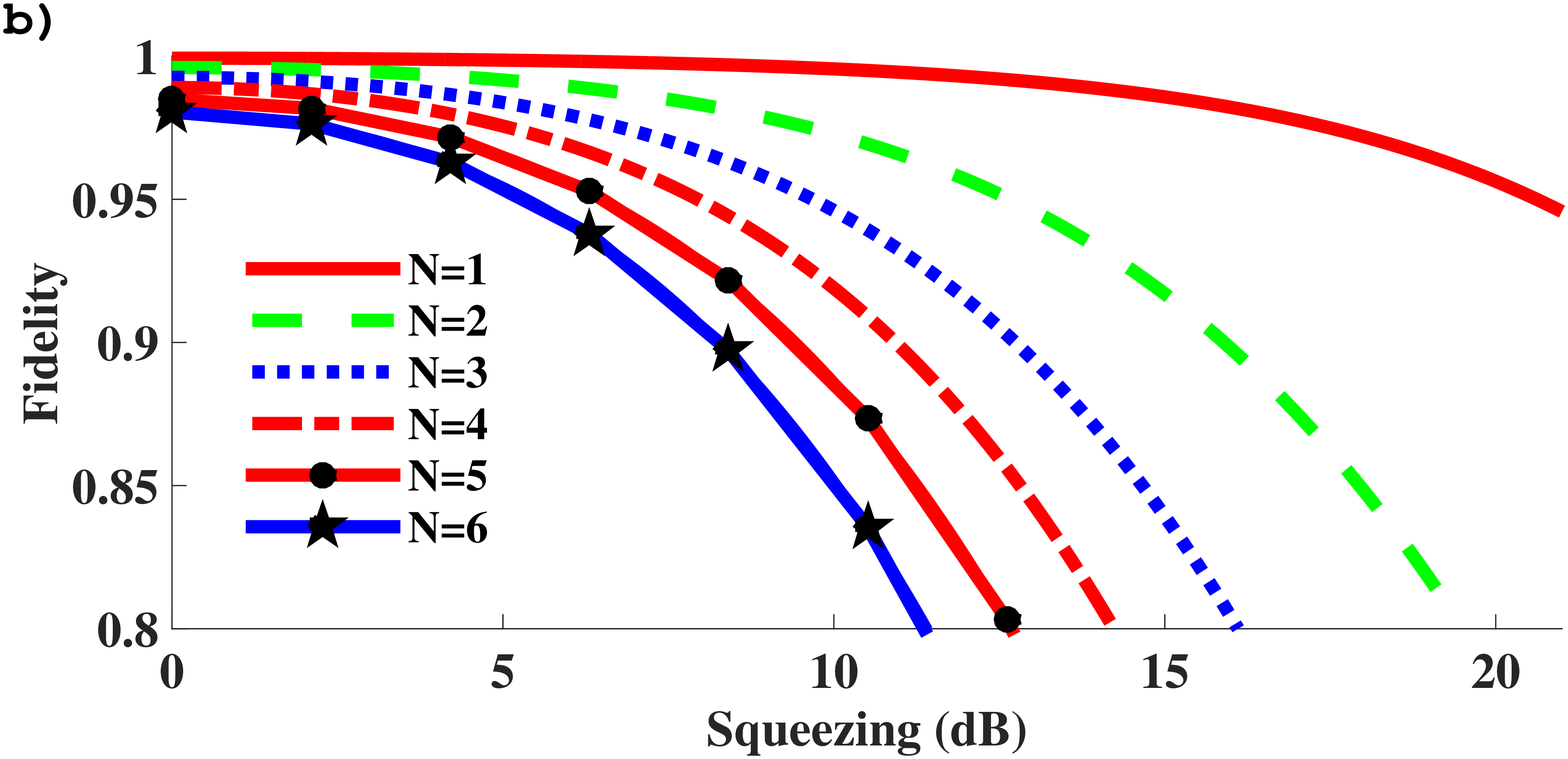}
\caption{\label{fig:fidelity-exp}(Color online) Fidelity as a function of the target squeezing between a linear graph state of $N$ nodes and the state generated using the protocol described in the text. Each data point is taken for an optimal evolution time (see text).  Each mechanical oscillator has a frequency $\Omega_j/ 2\pi= 11j~\mathrm{MHz}$ ($j=1,\ldots,N$, with $N=1,\dots,6$) and mechanical damping $\gamma/2\pi=32~\mathrm{Hz}$. The cavity mode decays with rate $\kappa/2\pi=0.2~\mathrm{MHz}$. The bath temperature for panel (a) is $T=15~\mathrm{mK}$ and for panel (b) is $T=1~\mathrm{mK}$}
\end{figure}

Given the findings of Sec.~\ref{sec:mech-noise}, our generation protocol performs better in the resolved-sideband regime and for high quality-factor oscillators at low temperature. High fidelity can then be achieved when $\gamma\ll\kappa\ll\Omega$ and $T\ll1$, a regime that has been extensively considered for sideband cooling in quantum electromechanical systems. For example, using experimental parameters of the order of the ones of Ref.~\cite{teufel:2011}, the values of fidelity shown in \reffig{fig:fidelity-exp} could be achieved.

Finally, let us briefly mention a possible readout strategy suitable to the setting here considered. In general, various approaches have been put forward in order to reconstruct the state of a single mechanical oscillator in an optomechanical setting (see Ref.~\cite{vanner2015towards} for a recent review). The generalization of those approaches to many mechanical oscillators could follow the path already pursued in other experimental contexts, such as ions or circuit-QED systems. In particular, in Ref.~\cite{tufarelli2012reconstructing} the state of a network of interacting oscillators is reconstructed via coupling only one of them to a two-level system, which plays the role of a probe that is eventually measured. The optomechanical setting here considered does not include any two-level system, nonetheless the scheme of Ref.~\cite{tufarelli2012reconstructing} could be adapted to the present case as well --- where now the radiation, rather than a two-level system, could act as a probe. In fact, a tomographic scheme along these lines is currently under study \cite{moore:in-preparation}. In particular, it is possible to show that such a scheme works both for the case of interacting and non-interacting mechanical oscillators in an optomechanical setting (\ie, both in the case of Fig.\ref{fig:optomech-system} and Fig.\ref{fig:optomech-system2}).
	
\section{Conclusions}\label{sec:conclusions}
	We have shown how to prepare arbitrary graph states of mechanical oscillators in the optomechanical system of Fig.\ref{fig:optomech-system}. The preparation is achieved by properly driving with external fields each of the two sidebands corresponding to each mechanical oscillator. The target graph state is generated by switching on and off a sequence of linearized Hamiltonians, which in turn can be achieved by changing the intensities and phases of the driving fields. At each switching step, the external pulses are applied for a sufficient time until a steady state is obtained by virtue of the optical losses.
	
In addition, we have considered the effect of non-zero mechanical noise, confirming the robustness of the present scheme. For a low number of mechanical oscillators and moderate squeezing, our protocol appears to be within reach of current technology since it requires linearized radiation-pressure interaction, resolved sideband, and low mechanical noise --- a regime that has been already achieved for the case of a single mechanical oscillator \cite{teufel:2011} and approached for the case of few oscillators \cite{lin2010coherent, massel2012multimode, shkarin2014optically}. In addition, experiments have been recently reported \cite{wollman2015quantum, pirkkalainen2015squeezing} where single-mode mechanical squeezing is achieved using the approach proposed in Ref.~\cite{clerk2013squeezing}. This is, in turn, very promising for the realization of the state generation protocol here presented, given that the latter extends and adapts the approach of Ref.~\cite{clerk2013squeezing} to generic multi-mode graph states. This work therefore identifies a promising path towards cluster states generation in mechanical systems, thus representing a first step towards measurement-based computation over continuous variables in a solid-state platform, rather then in the common optical setting.


\begin{acknowledgments}
	The authors would like to thank M.~Paternostro, S.~Pigeon, F.~Francica, and J.~Wragg for helpful insights. AF acknowledges funding from the John Templeton Foundation (Grant No. 43467). OH is supported by the Algerian ministry of higher education and scientific research (235/PNE/ENS/GB/2014-2015).

\end{acknowledgments}


\appendix


\section{Existence and uniqueness of Gaussian steady states}\label{sec:appendix-uniqueness}

	Consider an open system of $N$ modes with Hamiltonian $H$ and dissipation channels $L_1,\ L_2,\ldots,\ L_M$. The system evolves in time according to the master equation:
	\begin{equation}\label{eqn:master-equation-appendix}
		\dot{\rho}=-i[H,\rho]+\sum_k^M\left(L_k\rho L_k^\dagger-1/2L_k^\dagger L_k\rho-1/2\rho L_k^\dagger L_k\right)\ .
	\end{equation}
	We define $H=1/2\ R^T G R$ and $L=(L_1,\ L_2,\ldots,\ L_M)^T=C R$ with $G=G^T\in\mathds{R}^{2N\times 2N},\ C\in\mathds{C}^{M\times 2N}$ and $R=(q_1,\ q_2,\ldots,\ q_N,\ p_1,\ p_2,\ldots,\ p_N)^T$.

	For a Gaussian state of covariance matrix $V$ and mean value $\langle R\rangle$, the master equation~(\ref{eqn:master-equation-appendix}) is equivalent to \cite{Yamamoto:12}:
	\begin{eqnarray}
		\frac{\diff V}{\diff t}		&=&		A V+V A^T+B\label{eqn:covmat-diff}\\
		\frac{\diff \langle R\rangle}{\diff t}	&=&		A\langle R\rangle
	\end{eqnarray}
	where the matrices $A$ and $B$ are given by:
	\begin{eqnarray}
		A	&=&		\Sigma\left[G+\Im(C^\dagger C)\right]\\
		B	&=&		\Sigma\Re(C^\dagger C)\Sigma^T
	\end{eqnarray}
	and
	$$\Sigma=\left(\begin{array}{c@{\quad}c}0^{N\times N} & \mathds{1}^{N\times N}\\-\mathds{1}^{N\times N} & 0^{N\times N}\end{array}\right)\ ,$$
	where $\mathds{1}^{N\times N}$ and $0^{N\times N}$ are the identity and zero matrices respectively, and the superscript denotes the dimension of the matrix. Equation~(\ref{eqn:master-equation-appendix}) has a unique steady Gaussian state if and only if $A$ is Hurwitz matrix i.e., real part of each eigenvalue is negative.

	For the system described in \refsec{sec:system} given by the vector $R=(q_1,\ldots,q_N,q_{\mathrm{cavity}},p_1,\ldots,p_N,p_{\mathrm{cavity}})^T$, the Hamiltonian is given by equation~(\ref{eqn:hamiltonian}), and the matrices $G$ and $C$ are given in the block matrix form:
	\begin{eqnarray}
		G &=&
		\left(
		\begin{array}{l@{\quad}l@{\quad}l@{\quad}l}
			0^{N\times N}		&	\mathcal{A}^{N\times 1}		&	0^{N\times N}		&	\mathcal{C}^{N\times 1}\\
			(\mathcal{A}^{N\times 1})^T	&	0				&	\mathcal{D}^{1\times N}		&	0\\
			0^{N\times N}		&	(\mathcal{D}^{1\times N})^T	&	0^{N\times N}		&	\mathcal{B}^{N\times 1}\\
			(\mathcal{C}^{N\times 1})^T	&	0				&	(\mathcal{B}^{N\times 1})^T	&	0
		\end{array}
		\right)\\
		C	&=&		\sqrt{\frac{\kappa}{2}}\left(0^{1\times N}\quad,\quad 1\quad,\quad 0^{1\times N}\quad,\quad i\right)
	\end{eqnarray}
	where, as before, the superscript denotes the dimension of the corresponding matrix, and the matrix elements are found to be as follows:
	\begin{eqnarray}
		\mathcal{A}_j^{N\times 1}	&=&		g_j(\alpha_j^+\cos\phi_j^++\alpha_j^-\cos\phi_j^-)\\
		\mathcal{B}_j^{N\times 1}	&=&		g_j(-\alpha_j^+\cos\phi_j^++\alpha_j^-\cos\phi_j^-)\\
		\mathcal{C}_j^{N\times 1}	&=&		g_j(\alpha_j^+\sin\phi_j^++\alpha_j^-\sin\phi_j^-)\\
		\mathcal{D}_j^{1\times N}	&=&		g_j(\alpha_j^+\sin\phi_j^+-\alpha_j^-\sin\phi_j^-)
	\end{eqnarray}
	The eigenvalues of the matrix $A$ are:
	$$\underbrace{-\frac{\kappa}{4}\pm\sqrt{\left(\frac{\kappa}{4}\right)^2-\mathcal{A}^T\cdot \mathcal{B}+\mathcal{D}\cdot\mathcal{C}}}_{2\mbox{ times degenerate}}\quad,\underbrace{0\ ,\ldots,\ 0}_{2(N-1)\mbox{ times degenerate}}$$
	It is clear that when $N>1$, the matrix $A$ is not Hurwitz and therefore there is no unique Gaussian steady state for the system.

	For $N=1$, we have:
	$$\lambda^\pm=-\frac{\kappa}{4}\pm\sqrt{\left(\frac{\kappa}{4}\right)^2+g^2\left[(\alpha^+)^2-(\alpha^-)^2\right]}$$
	where $\alpha^\pm$ are the amplitudes of the two driving lasers.

	There exists a unique steady state if and only if $\alpha^+<\alpha^-$, and it is obtained after a time of the order of $\tau=\frac{1}{\Re{\lambda^+}}$ (see Appendix~\ref{sec:appendix-speed-steady-state}).


\section{Time scale to approach the steady state}\label{sec:appendix-speed-steady-state}
	The solution of the differential equation~(\ref{eqn:covmat-diff}) is:
	\begin{equation}\label{eqn:covmat-solution}
		V(t)=\EXP{A(t-t_0)}V_0\EXP{A^T(t-t_0)}+\int_{t_0}^t\diff s\ \EXP{A(t-s)}B\EXP{A^T(t-s)}
	\end{equation}
	where $V_0$ is the state of the system at time $t_0$. Diagonalizing the matrix $A=PDP^{-1}$, with $P$ given by the eigenvectors of $A$ and $D=\rm{Diag}(\lambda_1,\lambda_2,\ldots)$ by its eigenvalues $\lambda_j$, \refeq{eqn:covmat-solution} becomes:
	\begin{widetext}
	\begin{equation}\label{eqn:covmat-solution-2}
		V(t)=P\EXP{D(t-t_0)}P^{-1}V_0{P^T}^{-1}\EXP{D(t-t_0)}P^T+\int_{t_0}^t\diff s\ P\EXP{D(t-s)}P^{-1}B{P^T}^{-1}\EXP{D(t-s)}P^T
	\end{equation}
	\end{widetext}
	At the $k^{th}$ step of the switching scheme described in \refsec{sec:cluster-state}, we have the Hamiltonian~(\ref{eqn:hamiltonian-step-k}) and dissipator~(\ref{eqn:cavity-dissipation}). Hence the matrix $A$ has eigenvalues:
	\begin{equation}\label{eqn:eigenvals}
		\lambda^\pm=-\frac{\kappa}{4}\pm\sqrt{\left(\frac{\kappa}{4}\right)^2-\beta^2(1-r^2)}
	\end{equation}
with negative real part, implying a unique steady state. The speed at which the system approaches its steady state depends on how much negative is the real part of the eigenvalues $\lambda^\pm$: the more negative the faster the approach. Defining $\tau$ as the time scale to reach the steady state, we can estimate it as:
	\begin{equation}\label{eqn:tau}
		\tau=\frac{1}{\Re{\lambda^+}}
	\end{equation}
	and the condition of a maximum speed for approaching the steady state is given by:
	\begin{equation}
		\frac{\kappa}{4}\le\beta\sqrt{1-r^2}\;.
	\end{equation}
Under this condition, the shortest time scale to reach the steady state is
	\begin{equation}
		\tau_\mathrm{min}=\frac{4}{k}\ .
	\end{equation}
We should mention that, in order to prepare an infinitely squeezed state, an infinite amount of time is required, and this is clear from \refeq{eqn:tau} where the denominator vanishes.

	\bibliography{references.bib}

\end{document}